\documentclass[prd,aps,twocolumn, nofootinbib]{revtex4-1}
\usepackage[colorlinks=true,urlcolor=blue,linkcolor=,citecolor=blue]{hyperref}
\usepackage[utf8]{inputenc}
\usepackage{amsmath}
\usepackage{graphicx}
\usepackage{amssymb}
\usepackage{color}

\definecolor{verde}{rgb}{0,0.5,0}
\def\blu{\color{blue}}

\def\be{\begin{equation}}
\def\ee{\end{equation}}
\def\bea{\begin{eqnarray}}
\def\eea{\end{eqnarray}}
\def\be{\begin{equation}}
\def\ee{\end{equation}}
\def\ba{\begin{align}}
\def\ea{\end{align}}

\def\noi{\noindent}

\renewcommand\({\left(}
\renewcommand\){\right)}
\renewcommand\[{\left[}
\renewcommand\]{\right]}

\newcommand\lsim{\mathrel{\rlap{\lower4pt\hbox{\hskip0.5pt$\sim$}}
    \raise1pt\hbox{$<$}}}
\newcommand\gsim{\mathrel{\rlap{\lower4pt\hbox{\hskip0.5pt$\sim$}}
    \raise1pt\hbox{$>$}}}

\definecolor{purple}{cmyk}{0,.57,.22,0}

\begin{document}

%
%
%
\renewcommand{\topfraction}{0.99}
\renewcommand{\bottomfraction}{0.99}

 \title{Multimessenger Cosmology: correlating CMB and SGWB measurements}
\author{Peter Adshead,$^a$ Niayesh Afshordi,$^b$ Emanuela Dimastrogiovanni,$^c$ Matteo Fasiello,$^{d,e}$ Eugene A. Lim,$^f$ and Gianmassimo Tasinato$^g$\\ \vskip-0.1cm{\color{white}.}}

\affiliation{$^a$ Department of Physics, University of Illinois at Urbana-Champaign, Urbana, IL 61801, USA}
\affiliation{$^{b_1}$Department of Physics and Astronomy, University of Waterloo, 200 University Ave W, Waterloo,
Canada\\ $^{b_2}$Waterloo Centre for Astrophysics, University of Waterloo, Waterloo, ON, N2L 3G1, Canada\\
 $^{b_3}$Perimeter Institute For Theoretical Physics, 31 Caroline St N, Waterloo, Canada}
\affiliation{$^c$School of Physics, The University of New South Wales, Sydney NSW 2052, Australia}
\affiliation{$^d$ Instituto de F\'{i}sica Te\'{o}rica UAM/CSIC, Calle Nicol\'{a}s Cabrera 13-15, Cantoblanco, 28049, Madrid, Spain}
\affiliation{$^e$Institute of Cosmology and Gravitation, University of Portsmouth, Portsmouth, PO1 3FX, U.K.}
\affiliation{$^f$Physics Department, Kings College London, Strand, London WC2R 2LS, U.K.}
\affiliation{$^g$Department of Physics, Swansea University, Swansea, SA2 8PP, U.K.}

\begin{abstract}

Characterizing the physical properties of the stochastic gravitational wave background (SGWB) is a key step towards identifying the nature of its possible origin. We focus our analysis on SGWB anisotropies. The existence of a non-trivial primordial scalar-tensor-tensor (STT) correlation in the squeezed configuration may be inferred from the effect that a long wavelength scalar mode has on the  gravitational wave power spectrum: an anisotropic contribution.  Crucially, such a contribution is correlated with temperature anisotropies in the cosmic microwave background (CMB). We show that, for inflationary models that generate suitably large STT non-Gaussianities, cross-correlating the CMB with the stochastic background of gravitational waves  is a  very effective probe of early universe physics. The resulting signal can be a smoking-gun for primordial SGWB anisotropies.

\end{abstract}

\maketitle
\noindent
\noindent

\section {Introduction}
\label{intro}
 The recent detection \cite{Abbott:2016blz,TheLIGOScientific:2017qsa} of gravitational waves (GW) has ushered in a new era for GW astronomy. Operational and upcoming probes of the gravitational signal also hold the potential for transformative changes  in cosmology. From inflation  to (p)re-heating \cite{Grishchuk:1974ny, Easther:2006gt, GarciaBellido:2007dg, Dufaux:2007pt}, from cosmic strings to phase transitions \cite{Abbott:2017mem,Caprini:2019egz}, there is a plethora of early universe sources of GW that may be within reach of LISA \cite{Audley:2017drz}, DECIGO \cite{Kawamura:2006up}, the Einstein Telescope \cite{Maggiore:2019uih} and other 
 future missions \cite{Crowder:2005nr}. In this context, one of the key challenges will be to distinguish  primordial sources from astrophysical ones. Therefore, delivering a comprehensive  characterization of the stochastic gravitational wave background (SGWB) signal is paramount. Observables include the GW amplitude, scale-dependence, polarization, anisotropies \cite{Cusin:2017fwz,Jenkins:2018nty,Geller:2018mwu} and non-Gaussianities \cite{Adshead:2009bz}. In this work, we stress the importance of yet another handle on GW cosmology, namely cross-correlations with   probes of the electromagnetic spectrum. We describe under what conditions the cross-correlation of CMB with SGWB data may reveal precious information on early universe physics. 
 
Inflation  generates three-point correlation functions of the scalar-tensor-tensor (STT) type. In the limit of a long-wavelength (soft) scalar mode, this  STT correlator may be probed by studying the effect of the long mode on the GW power spectrum. The effect, as we demonstrate in this paper, takes the form of an anisotropic GW signal. Because the GW anisotropies are generated by the same scalar density fluctuations, this anisotropy is correlated with the cosmic microwave background anisotropies.   This cross-correlation is not only sensitive to primordially-induced GW anisotropies, but also to those due to propagation of GW through an inhomogeneous Universe (see e.g.\ \cite{Alba:2015cms,Contaldi:2016koz,Bartolo:2019oiq,Bartolo:2019yeu}). However, while the propagation anisotropies are necessarily small, we show that in the presence of a primordial (squeezed) bispectrum component, the cross-correlation can be very large. 

Several classes of inflationary models, those generating a sufficiently large STT-type non-Gaussianity, may deliver the leading contribution to the GW anisotropies and can therefore be tested also via cross-correlation. All such setups go beyond the minimal single-field slow-roll paradigm. Multi-field models are the prototype of well-motivated \cite{Baumann:2014nda} non-minimal inflationary realizations that may generate large bispectra in the soft limit.  Non-Bunch-Davies vacua exemplify the possibility of excited initial states whose signatures include a different momentum dependence of the three-point function with respect to the standard scenario.  Finally, models with alternative symmetry-breaking patterns exhibit a rich and distinctive phenomenology.

This paper is organized as follows: we begin in Section \ref{sec2} with a review of the origin of \textit{intrinsic} (a synonym of ``primordial'' for the purposes of our work) and induced anisotropies of the gravitational wave power spectrum; in Section \ref{sec4}, we cross-correlate the GW anisotropy with those in the cosmic microwave background and compute the corresponding statistical uncertainty; we devote section \ref{sec5} to a survey of inflationary models that can support a large STT correlator in the squeezed configuration; %
we conclude in Section \ref{sec6}.  

\section{ SGWB anisotropies}
\label{sec2}

 In this section, we show how primordial scalar curvature fluctuations,  $\zeta$, induce an anisotropic contribution, $\delta_{\rm GW}$, to  the energy density associated with the SGWB. We begin by reviewing anisotropic SGWBs to introduce our notation (see, e.g., Ref.\ \cite{Maggiore:1999vm}), before focussing on intrinsic and induced sources of anisotropy in Sections \ref{sec:intrinsic} and \ref{sec3}, respectively.
 
Experimental searches for a SGWB signal focus on the normalized energy density per unit frequency
 \be
 \Omega_{\rm GW} (f, {\bf x})\,\equiv\,\frac{1}{\rho_{cr}}\,\frac{d\rho_{\rm GW}}{d\ln f},
 \ee
 where $\rho_{cr}$ is the critical density,  $\rho_{\rm GW}$ is the GW energy density. We have allowed for position dependence via ${\bf x}$, the magnitude $|{\bf x}|$ accounts for the time elapsed from GW horizon (re-)entry until today, while $\hat{x}$  indicates the direction of observation. The quantity  $\Omega_{\rm GW} (f, {\bf x})$ is obtained by averaging the GW signal over all GW directions $\hat n$
\bea \label{omega}
\Omega_{\rm GW} (f , {\bf x})&\equiv&
\frac{1}{4 \pi}\, \int d^2 \hat  n \,\omega_{\rm GW}(f,\,\hat n, {\bf x}) \; .
\eea

One codifies within $\omega_{GW}(f,\,\hat n, {\bf x})$ both an isotropic part as well as possible SGWB anisotropies. The parametrization is  (see also e.g. \cite{Alba:2015cms,Contaldi:2016koz,Bartolo:2019oiq,Bartolo:2019yeu}) as follows:
\be \label{omsmall}
\omega_{\rm GW}(f,\,\hat n, {\bf x})\,=\,\bar \omega_{\rm GW}(f)\,\left[ 1+ \delta_{\rm GW} (f,\,\hat n, {\bf x})\right]\;~.
\ee
In general, there are two distinct effects that source the anisotropy $\delta_{\rm GW}$: intrinsic or primordial anisotropies, and induced anisotropies. We  now consider each case separately. We note here that the standard sub-horizon evolution is implicit in Eqs.~(\ref{omega}-\ref{omsmall}). More specifically, in Eq.~(\ref{omsmall}) it is the factored out term $\bar \omega_{\rm GW}(f)$ to be understood as the result of the usual time evolution (all the way to today) regulated by the appropriate transfer function.

\subsection{Intrinsic SGWB anisotropies}\label{sec:intrinsic}

 It is well-known that large non-Gaussianities can induce anisotropies in the SGWB. A case in point is the analysis of Ref.\ \cite{Ricciardone:2017kre,Dimastrogiovanni:2019bfl} showing how a large squeezed  tensor 3-point  function leads to a quadrupolar asymmetry  in the SGWB. In the present work, we consider instead the STT case arising from a \textit{primordial} $\langle \zeta \gamma \gamma  \rangle$ correlator. 
   
We work in comoving gauge, where  the perturbed Friedmann-Robertson-Walker line element reads
\bea
{\rm d}s^2=-a^2(\eta)\left\{{\rm d}\eta^2 - \[(1-2\zeta)\delta_{ij} + \gamma_{ij}\] {\rm d}x^i {\rm d}x^j\right\}\; ,
\eea
where, $\eta$ is conformal time,  $\zeta$ is the scalar curvature fluctuation, and $\gamma_{ij}$ is the transverse-traceless  tensor perturbation. The scalar and tensor power spectra are
\bea\label{defpow}
\langle\zeta_{{\bf k}_1}\,\zeta_{{\bf k}_2} \rangle'\,\equiv\,P_\zeta \left({ k}_1\right)\,,\qquad
\langle\gamma_{{\bf k}_1}\,\gamma_{{\bf k}_2} \rangle'\,\equiv\,P_\gamma \left({ k}_1\right)\;,
\eea
and the scalar-tensor-tensor correlator bispectrum $B$ is defined as:
 \be\label{defbis}
B_{STT}\left({\bf  k}_1,\,{\bf  k}_2,\,{\bf q}_L\right) \equiv  \langle\gamma_{{\bf k}_1}\,\gamma_{{\bf k}_2}\, \zeta_{{\bf q}_L} \rangle' \; ,
 \ee
where the symbol $'$ denotes  the correlators, excluding the wavenumber delta functions.
 
In analogy\footnote{The pioneering work on fossil fields includes \cite{Jeong:2012df,Dai:2013kra,Brahma:2013rua,Dimastrogiovanni:2014ina,Dimastrogiovanni:2015pla}.} with the results of \cite{Dimastrogiovanni:2019bfl}, we find that the existence of mode-coupling, in the form of a non-trivial squeezed bispectrum, modulates the primordial tensor power spectrum according to 
  \begin{equation}\label{pgammod}
 P_{\gamma}^{\text{mod}}(\textbf{x},\textbf{k}_{})=P_{\gamma}^{}(k_{})\left[1+\int\frac{d^{3}q_{L}}{(2\pi)^{3}}\,e^{i\textbf{q}_{L}\cdot\textbf{x}}\,F_{\rm NL}(\textbf{q}_{L},\textbf{k}_{})\,\zeta(\textbf{q}_{L})\],
 \end{equation}
where
 \bea
 F_{\rm NL}\left(
 {\bf q}_L,\,
 {\bf k}
 \right)\,=\,\frac{B_{STT}\left({\bf k}+ {\bf q}_L/2,\,-{\bf k}+ {\bf q}_L/2,\,- {\bf q}_L\right)}{ P_\gamma ({ k})\, P_\zeta ({ q_L})}.\;\;\;\;\;
 \label{bispe}
 \eea
The modulated spectrum in Eq.~(\ref{pgammod}) is to be understood as a primordial quantity, i.e. the spectrum at horizon re-entry for the $k$ modes. Its sub-horizon evolution
is accounted for in standard fashion through the transfer function in the expression for the energy density $\Omega_{\rm GW}$.
 
We stress here the implicit assumption that the STT bispectrum in Eq.~(\ref{bispe}) is one that breaks consistency relations (CRs). If this were not the case, the leading order term in the squeezed $B_{\rm STT}$ would amount to a gauge artifact and the resulting physical contribution would be suppressed (see Section \ref{sec5} for examples of models that break CRs). 

We further note that the integral in Eq.~\eqref{pgammod} only spans large scales ($q_L\ll k$). Expressing quantities in terms of GW frequency and direction, the resulting GW energy density is
 \bea
\Omega_{\rm GW} (f,\textbf{x})
=\frac{2\,\pi^2}{3\,H_0^2}\,{f^2}\,
\Bigg[\frac{1}{4 \pi}\, \int d^2 \hat  n \,P_\gamma^{\rm mod} (f,\,\hat n,\textbf{x})\ \Bigg], \;\;\;\; 
\eea
and the quantities $\bar \omega_{GW}$, $\delta_{\rm GW}$ introduced in Eq.~\eqref{omsmall} read
\bea
\bar \omega_{\rm GW}(f)=\frac{2\,\pi^2}{3\,H_0^2}\,{f^2}\,P_\gamma ({ f})\; ,
\eea
\bea \label{def33}
&&\delta_{\rm GW}^{\rm prim} (f,\,\hat n,\textbf{x})= \qquad  \\ &&\int_{|f_L|\ll | f|} f_L^2\,d f_L \int d^2 \hat n_L\,e^{i\textbf{q}_{L}\cdot\textbf{x}} \,F_{\rm NL}\left(f_L  \hat n_L, \,f \hat n \right)\,\zeta\left(f_L\,\hat n_L\right) \; ,  \nonumber
\eea
where ${\bf q}_L=2\pi f_L\hat n_L$ and ${\bf k} = 2\pi f \hat n$. The intrinsic anisotropy can therefore be very significant in models with an enhanced STT correlator. In such cases the primordial contribution  becomes the leading one and the induced counterpart, $\delta_{\rm GW}^{\rm ind}$, to be described in the next Section, can be safely neglected.

\subsection{Induced SGWB anisotropies}
\label{sec3}

A second source of anisotropy of the SGWB is due to the scalar perturbations present in an inhomogeneous universe as primordial GWs travel over cosmological distances. The effect is the GW analog of the Sachs-Wolfe effect: long wavelength scalar fluctuations
modulate the GWs frequency and direction and induce anisotropies in $\Omega_{\rm GW}$ (see e.g. \cite{Alba:2015cms,Contaldi:2016koz,Bartolo:2019oiq,Bartolo:2019yeu}). We consider the case of scalar fluctuations that re-enter the horizon  during matter domination.\footnote{Our conclusions do not depend qualitatively on this assumption. See e.g. \cite{Bartolo:2019yeu} for the general case.} Following \cite{Alba:2015cms}, we assume the gravitational waves reenter the horizon when the universe can be treated as a perfect fluid with $p=w \rho$. The induced SGWB anisotropy (see \cite{Alba:2015cms,Bartolo:2019yeu} for a derivation) is given by
\bea
\label{definani}
\delta_{GW}^{\rm ind} (f,\,\hat n)&=&\frac25\,\left(\nu+1 - \frac{n_T}{2}\right)\zeta_L\left(\hat n\right)\; ,
\eea
with $ \nu=2/(1+3 w)$ and $n_T$ the standard tensor tilt. The overall coefficient on the right hand side of Eq.~(\ref{definani}) is then expected to be of order one. This is to be compared with Eq.~(\ref{def33}), whose contribution can be significantly larger than the induced term for large primordial STT non-Gaussianity.  Indeed, inflationary models such as those in Section \ref{sec5} can support signals  in excess of  $F_{\rm NL}\sim 10^3$. 

In what follows we consider cross-correlation of GW anisotropies with CMB temperature anisotropies under the assumption that the former receives the leading contribution via Eq.~(\ref{def33}).\footnote{Curvature fluctuations also contribute to anisotropies of the astrophysical SGWB: $\zeta$ induces matter inhomogeneities in the large scale structure, where astrophysical GW sources are located. The amplitude of the resulting SGWB anisotropies are at most of the order of $\delta_{GW}^{\text{ind}}$, see e.g. \cite{Cusin:2017fwz}.}

\section{Correlating anisotropies in the SGWB with the CMB}
\label{sec4}

\begin{figure}[t]
	\centering
	\includegraphics[width=0.39\textwidth]{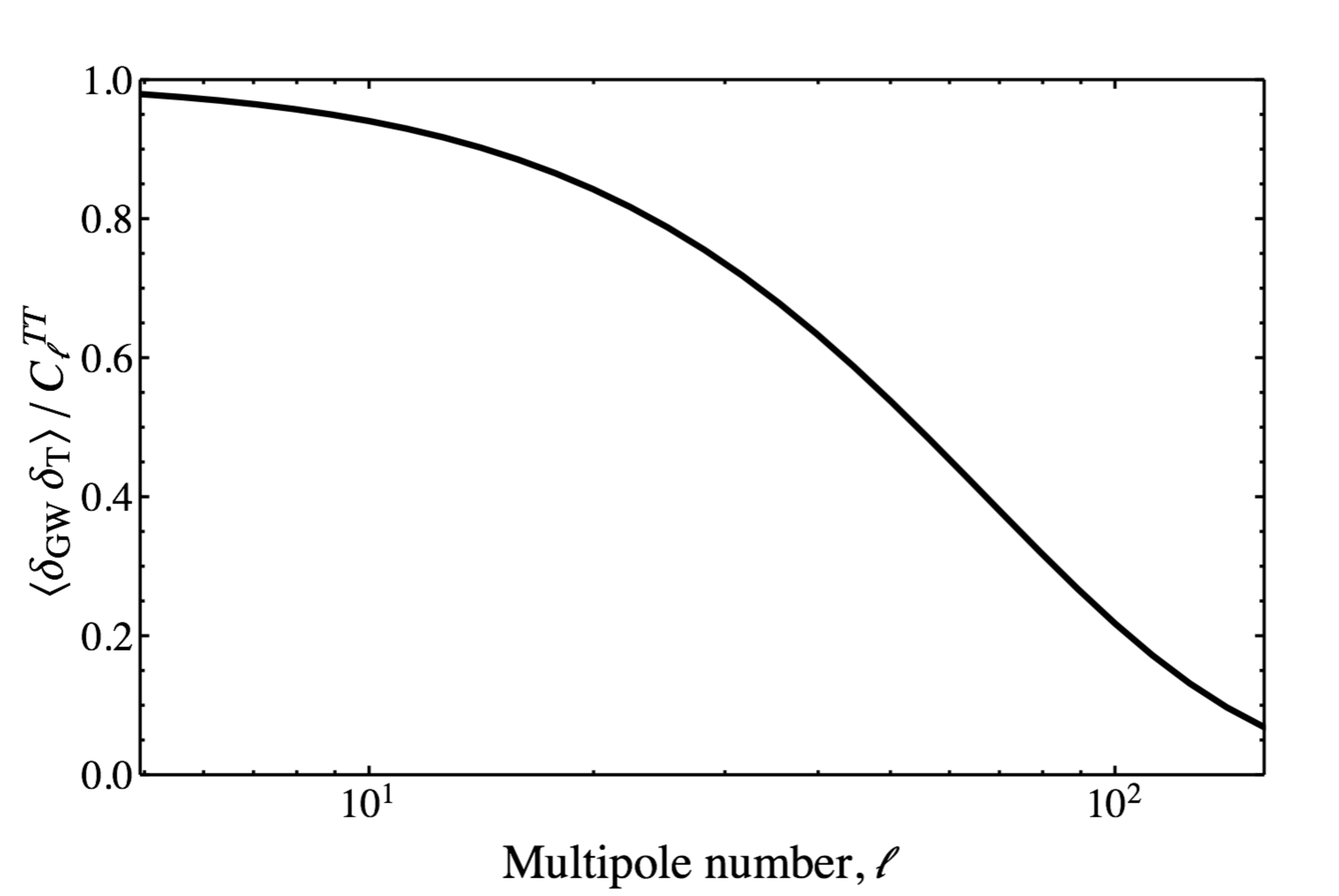}
	\caption{Cross-correlation of the GW anisotropy with the CMB in the Sachs-Wolfe limit. The correlation decays exponentially with a characteristic scale $\ell_* \approx 44$.}
	\label{fig6}
\end{figure}

We now compute the cross-correlations between the gravitational wave anisotropies and the CMB-temperature anisotropy, focusing on the contribution due to $\delta_{GW}^{\text{prim}}$. The starting point is given by 
\begin{equation}\label{nuov4}
\delta_{GW,\ell m}^{\text{prim}}=\int d\Omega_{\hat{x}}\,Y^{*}_{\ell m}(\hat{x})\int_{k_{}\ll k_{*}}\frac{d^{3}k_{}}{(2\pi)^{3}}\,e^{i\textbf{k}_{}\cdot\textbf{x}}\,F_{\rm NL}(k_{},k_{*})\,\zeta(\textbf{k}_{})\, ,
\end{equation} 
where $Y^{*}_{\ell m}(\hat{x})$ are spin-0 spherical harmonics, 
$k_{*} = 2\pi f_{*}$, is the short modes momentum  and, for the sake of simplicity, we consider a $F_{\rm NL}$ that depends only on the magnitude of the momenta in Eq.~(\ref{bispe}). Similarly, for temperature anisotropies one has:
\begin{equation}\label{nuov101}
\delta_{T,\ell^{}m}=(-i)^{\ell}\,\frac{4\pi}{5}\,\int\frac{d^{3}k}{(2\pi)^{3}}\,Y^{*}_{\ell^{} m}(\hat{k})j_{\ell}(k\, r_{lss})\zeta(\textbf{k})\,,
\end{equation}
where the subscript ``$_{lss}$'' in $r_{lss}\,=\,\eta_0-\eta_{lss}$ signals the last scattering surface.
The resulting angular cross-power spectrum is 
\begin{align}\label{nuov46}
&\langle \delta_{GW,\ell m}^{\text{prim}}\,\delta_{T,\ell^{'} m'}^{*}\rangle\equiv \delta_{\ell\ell^{'}}\delta_{mm'} C^{GW-T}_{\ell}= \\ &\delta_{\ell\ell^{'}}\delta_{mm'}
\cdot\frac{2}{5\pi}
\int dk\,k^{2}j_{\ell^{}}(k\,r_{lss})j_{\ell^{}}(k\,r_{*})F_{\rm NL}(k_{},k_{*})P_{\zeta}(k)\,, \nonumber
\end{align}
where $r_{*}$ is the time between between today and horizon re-entry for the tensor modes.  The expression in Eq.~(\ref{nuov46}) can be further simplified under the  assumptions of (i) a scale invariant power spectrum $P_{\zeta}(k)=(2\pi^{2}/k^{3})A_{s}$; (ii) a constant value for $F_{\rm NL}$. The corresponding $\ell$ dependence of $C^{GW-T}_{\ell}$ is represented in Fig.~(\ref{fig6}), normalised by $C^{T-T}_{\ell}$, for the case of $F_{\rm NL}=1$. The plot is largely insensitive to the value of $r_{*}$ so long as this quantity corresponds to modes that have re-entered the horizon during radiation era, i.e. those that are relevant both for PTA and laser interferometers such as LISA.

It is both useful and straightforward \cite{Verde:2009tu}  at this stage to provide an estimate of the statistical error in the measurement of the key primordial quantity $F_{\rm NL}$.
Equipped with the entries of the matrix 

\begin{equation}
	\textbf{C}_{\ell} = 
	\begin{pmatrix}
		C_{\ell}^{TT} & C_{\ell}^{T-GW}  \\
		C_{\ell}^{T-GW} & C_{\ell}^{GW} 
	\end{pmatrix}\,,
\end{equation}
where $C_{\ell}^{TT}=2\pi\,A_{s}/[25\,\ell(\ell+1)]$ (in the Sachs-Wolfe regime), and $C_{\ell}^{GW}$ is the auto-correlation of GW anisotropies, we obtain the Fisher matrix for $F_{\rm NL}$
\bea
F\simeq \sum^{\ell_{\rm max}}_{\ell=\ell_{\rm min}}\frac{(2\ell +1) A_s^2 I^2_{\ell}}{C^{TT}_{\ell} N^{GW}_{\ell}} \;.
\eea  
$I_{\ell}$ is given by   $I_{\ell}= (4 \pi/5)
\int dk\cdot 1/k\cdot\, j_{\ell^{}}(k\,r_{lss})j_{\ell^{}}(k\,r_{*})$ and
we have assumed the Sachs-Wolfe approximation and a noise-dominated regime, $C^{GW}_{\ell}\simeq N^{GW}_{\ell}$, for GW anisotropies.~The error on $F_{\rm NL}$ is simply given by $\delta F_{\rm NL}=1/\sqrt{F}$. In calculating $\delta F_{\rm NL}$ {we follow \cite{Alonso:2020rar} and adapt\footnote{We are indebted to Ameek Malhotra for precious help and insight on this matter.} the associated code schNell to the case of BBO. We refer the interested reader to \cite{Malhotra:2020ket} for more details.} We find that a relative error of a few percent for $\delta F_{\rm NL}/F_{\rm NL}$ is achievable by BBO if e.g. (i) $F_{\rm NL}\sim 10^3$ and $n_T=0.25$ or (ii) $F_{\rm NL}\sim 10^5$ and $n_T=0.12$, to give a few examples. Here $n_T$ is the spectral index of the primordial GW spectrum and we refer the reader to the next section \ref{sec5} for inflationary models that support a positive  $n_T$ and sizeable values for $F_{\rm NL}\,$.

\section{Survey of Inflationary Models}
\label{sec5}
 
The detection of a primordial non-Gaussian signal in the squeezed (soft) limit by current or near-future probes would be a tell-tale sign of an early universe scenario beyond the single-field slow-roll (henceforth ``minimal") inflationary paradigm. Several scenarios that go beyond the minimal realization break the so-called consistency relations (CRs). CRs connect the squeezed limit of an $N+1$-point function with its $N$-point function counterpart. They stem from a residual diffeomorphism (diff)  in the description of a physical system. CRs are famously \cite{Maldacena:2002vr} in place for minimal inflationary setups. As a result of CRs, the leading contribution to the soft bispectrum limit can be shown to be a gauge artifact (see e.g. \cite{Tanaka:2011aj,Pajer:2013ana}), thus leading to a suppressed physical signal.   
Crucially, whenever CRs are broken,  the leading three-point function contribution may be physical, which can lead to enhanced signals in the squeezed limit. 

Schematically, CRs are broken (i) whenever several long modes transform both non-linearly and independently under a residual gauge diff; (ii) when the vacuum is modified with respect to the ``minimal" Bunch-Davies prescription; or (iii) whenever the inflationary background breaks space diffs or, in general, has a different symmetry breaking pattern from standard single-clock inflation. In addition to CRs breaking, a feature for the models we are after is a blue-tilted tensor power spectrum, or more broadly one that is significant at small scales. This is already known to occur when isocurvature fields are present \cite{Iacconi:2019vgc} and alternative diff breakings considered \cite{Endlich:2012pz,Bartolo:2015qvr}.

We  now provide  an estimate of the size of the STT correlator for a general class of multi-field models that falls under category (i).  We then argue that it is reasonable to expect an enhanced signal also in classes of inflationary models belonging to category (ii) as well as (iii).

\subsection{Isocurvature fields}
Consider a setup comprising an extra spin-2 particle $\sigma_{ij}$ during inflation. We focus on the case of a $\sigma_{ij}$ directly (i.e. non-minimally) coupled to the inflaton field. This allows $\sigma_{ij}$ to be effectively light compared to the Hubble scale and, in turn, the bispectrum to have a significant squeezed component \cite{Bordin:2018pca,Dimastrogiovanni:2018gkl}. The mixing quadratic and cubic Lagrangian of the effective setup read 
\begin{align}
\label{action}
\mathcal{S} &\supseteq \;\; \, \int dt \, d^3x\, a^3 \Big[ -\frac{g}{\sqrt{2 \epsilon} H} a^{-2} \partial_i \partial_j \pi_c \sigma^{ij} +\frac{1}{2} g \dot{\gamma_c\,}_{ij} \sigma^{ij}  \Big]  \; \nonumber\\
&\;\;-\, \int dt \, d^3x\, a^3 \Big[ \frac{g}{2\epsilon H^2 M_{\rm Pl}}a^{-2} (\partial_i\pi_c \partial_j\pi_c \dot{\sigma}^{ij}\\ \nonumber&\qquad \qquad \qquad\qquad+2H \partial_i\pi_c \partial_j\pi_c {\sigma}^{ij} ) +\mu(\sigma^{ij})^3 +\dots \Big], \; 
\end{align}
where $m$ is the mass of the spin-$2$ field, $\gamma_c = \gamma M_{\rm Pl}$, and $\pi_c\equiv\sqrt{2 \epsilon_1} H M_{\rm Pl}\, \pi$ is the canonically normalized Goldstone boson, linearly related to the curvature fluctuation via $\zeta\simeq-H \pi$.  The quantities $g,\mu$ are coupling constants, and  $c_{i}$ labels the sound speeds for the helicity modes (0,1,2) of $\sigma_{ij}$.
\begin{figure}[t!]
	\centering
	\includegraphics[scale=0.25]{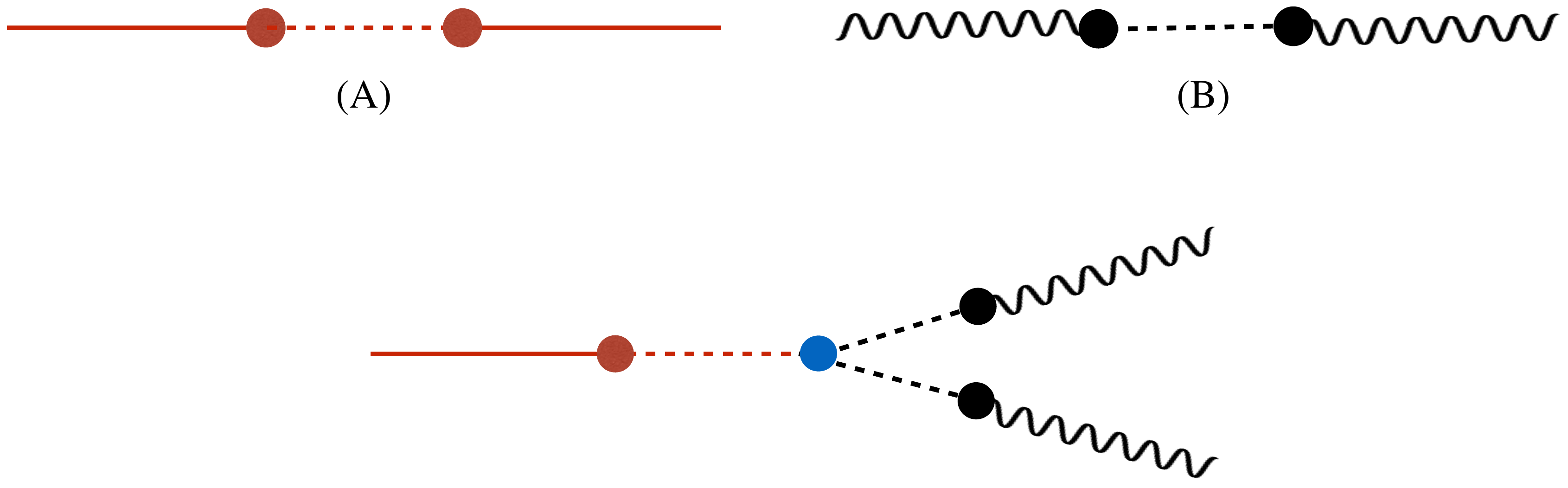}
	\caption{Top: $\sigma$-sourced contributions to the scalar (A) and tensor (B) power spectra. Bottom: main contribution to $\langle \zeta \gamma \gamma\rangle$. The solid line stands for $\zeta$ propagators, wiggly lines are for $\gamma$, dashed lines are for $\sigma$. The bispectrum diagram at the bottom relies on the $\mu\,\sigma^3$ interaction in the cubic Lagrangian (see 3rd line of Eq.~\ref{action}) for the vertex. It builds instead on the $\sigma_0 \pi$ and $\sigma_2 \gamma$ mixing in the quadratic Lagrangian (1st line of Eq.~\ref{action}) to contract external legs with internal ones.}
	\label{fig33}
\end{figure}
For completeness, we provide here the free part of the quadratic action for $\sigma$ (see \cite{Bordin:2018pca} for a derivation): 
\bea
&S_2[\sigma]=\frac{1}{4}\int dt dx^{3} a^3 \Big[(\dot{\sigma}^{ij})^2 -c_2^2 a^{-2}(\partial_i {\sigma}^{jk})^2 \qquad \qquad \qquad \qquad \nonumber \\ &\qquad\quad\quad-\frac{3}{2}(c_0^2 -c_2^2)a^{-2} (\partial_i {\sigma}^{ij})^2-m^2 ({\sigma}^{ij})^2  \Big]\,.
\eea 
The diagram at the bottom of Fig.~\ref{fig33} represents the leading-order contribution to $\langle \zeta \gamma \gamma \rangle$. We  focus on the $c_{2}\ll 1$ regime where the leading contribution to metric tensor modes is due to $\sigma_{ij}$ (Diagram B at the top of Fig.~\ref{fig33}). One can estimate, via the in-in formalism, the amplitude of the bispectrum contribution to obtain %
\bea
\langle \zeta \gamma \gamma \rangle\Big|_{\sigma-{\rm mediated}}\sim \Delta_{\zeta}^{1/2} \Delta_{\gamma} \left(\frac{g}{H \sqrt{\epsilon}} \right) \left(\frac{\mu}{H}\right)  \frac{1}{c_{2}^{2\nu}} \; ,
\label{est}
\eea
where, as in the setup of Ref.\ \cite{Bordin:2018pca}, we impose $\mu/H<1, \; g/(H\sqrt{\epsilon})<1$, and $\Delta^2_{\zeta, \gamma} = k^3 P_{\gamma, \zeta}(k)/2\pi^2$. The amplitude of the bispectrum then reads
\bea\label{fnl}
F_{\rm NL}= \frac{\langle \zeta \gamma \gamma \rangle}{P_{\zeta} P_{\gamma}} \sim \mathcal{O}\(\frac{\mu}{H}\)\times \frac{1}{c_2^{2\nu}} \frac{1}{\Delta^{1/2}_{\zeta}} \; .
\label{est_2}
\eea

For very light extra field content, a scaling of the form  $\sim 1/(k_1^3 k_3^3)$ is expected in the squeezed configuration of the bispectrum. Sizable STT non-Gaussianity is therefore possible by virtue of the terms $1/\Delta_{\zeta}$ and $1/c_2^{ 2 \nu}$, given that $c_2\ll1$. For a nearly massless extra spin-2 one has $\nu\sim 3/2$, which gives overall $F_{\rm NL}\propto 1/c_2^3$. Assuming conservatively that the small factor $\mathcal{O}(\mu/H)$  due to $\mu/H$ and $g/(H\sqrt{\epsilon})$ is counterbalanced by $\Delta^{-1/2}_{\zeta}$, a value of the sound speed\footnote{This range of values for $c_{2}$ is safe in terms of observational constraints, perturbativity, and stability \cite{Bordin:2018pca,Iacconi:2019vgc}.} in the range $1/100\leq c_2\leq 1/10$ corresponds to $F_{\rm NL}\simeq  10^3-10^6$. These numbers are not expected to qualitatively change upon performing the full calculation through the \textit{in-in} formalism. We leave this to future work. The enhanced bispectrum in Eq.~(\ref{est_2}) motivates our analysis of cosmological (cross-)correlations sensitive to a squeezed primordial signal. 
{It is worth stressing at this stage the following notion. Both a non-trivial GW power spectrum scale dependence (e.g. a blue tilt) and a sufficiently large STT non-Gaussianity are necessary for the signal we are after to be detectable. We shall point to recent literature \cite{Iacconi:2019vgc,Iacconi:2020yxn} that shows how, for example, the effective field theory description of isocurvature modes just discussed can satisfy both such conditions. The work in \cite{Iacconi:2020yxn} showed in particular how such set-up can generate a tensor tilt $n_T\sim 0.27$, certainly within reach for BBO, and a sufficiently large STT i.e. one in the $10^3 < F_{\rm NL} < 10^6$ range, whilst satisfying all available CMB bounds.}

 In order to provide a broader overview of models that may engender a large squeezed STT correlator, we now move on to setups with, respectively, excited initial states and alternative symmetry breaking patterns. Our analysis in what follows is more qualitative in nature compared to Subsection A; we leave a more in-depth study for future work \cite{future}.%
 
\subsection{Excited initial states}
The existence of a pre-inflationary era may be probed by exploring the signatures of excited initial states. In the most general case, both scalar and tensor fluctuations can have non-Bunch-Davies (nonBD) initial conditions.  \\
\noi NonBD initial states were first investigated for the  scalar degree of freedom in single-field slow-roll (SFSR) inflation \cite{Holman:2007na} and then extended to the effective field theory (EFT) framework \cite{Agarwal:2012mq}.  For derivative interactions such as those in the EFT (e.g. $\dot{\zeta}^3\; , \dot{\zeta}(\partial_i\zeta)^2$), the squeezed limit of $\langle \zeta^3\rangle$ has a $k_{\rm Short}/k_{\rm Long}$ enhancement  with respect to the so-called local shape.\footnote{The local template has a simple scaling $\sim 1 /(k_{\rm Long}^3\; k_{\rm Short}^3)$.} This is  particularly relevant for correlations between modes at widely different scales. 

{ 
We use the analogy with the scalar bispectrum  to estimate the $\langle \zeta\gamma\gamma\rangle$ momentum scaling when interactions beyond those of SFSR are considered. Let us focus on interactions present in the EFT approach to inflation \cite{Cheung:2007st}. One may recall that for $\langle \zeta^3\rangle$, the typical interactions in the EFT of inflation are $\dot{\zeta}^3\; , \dot{\zeta}(\partial_i\zeta)^2$. Because analogous interactions can be found in the 
cubic tensor and STT Lagrangians (see e.g., \cite{Naskar:2018rmu}), we expect in the EFT case a $k_{S}/k_{L}$ enhancement for $\langle \zeta \gamma \gamma\rangle$ similar to the one for $\langle \zeta^{3}\rangle|_{\rm EFT}$. 
Furthermore,  using an EFT framework,  Ref.\ \cite{Akama:2020jko} finds that tensor-scalar-scalar bispectra are amplified for nonBD initial states, which supports our expectation of a large squeezed $\langle \zeta\gamma\gamma\rangle$ signal.}

 \subsection{Alternative symmetry-breaking patterns}

In solid inflation \cite{Endlich:2012pz}, scalar fields with a space-dependent background break spatial translations and rotational invariance. The homogeneity and isotropy of the cosmological background is preserved by resorting to internal symmetries of the theory. The violation of consistency relations in solid inflation, including for $\langle \zeta\gamma\gamma\rangle$, has been verified \cite{Endlich:2013jia}, with a scaling of the squeezed limit bispectra similar to the local template. This result implies a strong enhancement with respect to SFSR inflation which can potentially be constrained by means of  CMB-interferometer cross-correlations studied in this paper. Similar considerations apply for supersolid inflation \cite{Bartolo:2015qvr, Ricciardone:2016lym}, a related scenario with non-standard symmetry breaking patterns. We conclude that this class of models can deliver a naturally large signal for mixed bispectra that can potentially be tested via CMB-interferometer cross-correlations.\\

It is worth to point out at this stage that it is in principle possible to distinguish among the various classes of models in section \ref{sec5} by focussing, for example, on the $k$ scaling of non-Gaussianities in the squeezed configuration. As mentioned above, in the case of excited initial states the three-point function is enhanced w.r.t.\ the standard single-field slow-roll scenario by integer powers of $k_{\rm S}/k_{\rm L}$. This is in contradistinction to the case discussed in subsection \ref{sec5} A whose three-point function momentum scaling is generally non integer (i.e. non analytical) in powers of $k_{\rm L}/k_{\rm S}$. The exponent $x$ in $(k_{\rm S}/k_{\rm L})^x$ will depend on both the mass and the spin of the iso-curvature field(s) \cite{Arkani-Hamed:2015bza}. As for models such as solid inflation (section \ref{sec5} C), the STT scaling is the same as the leading term in single-field slow-roll inflation (but the tensor spectrum is blue in solid inflation). It is therefore different from both the one for the class of models in section \ref{sec5} A (unless mass and spin of the iso-curvature mode combine in such a way as to precisely give an analytical scaling) and those in section \ref{sec5} B.

\section{Conclusions}
\label{sec6}

The past few years have witnessed an explosion of new   exciting opportunities for gravitational wave physics. The detection of signals from astrophysical sources has provided key tests of the models of stellar evolution, astrophysics, gravity, and dark energy. Ground based and, in the near future,  space-borne interferometers hold the potential to test the very early universe, including the inflationary era where many models produce GW signals that can be detected by e.g., LISA observatory. In this work, we have investigated how to further characterize the primordial component of the stochastic background of GW. The physics corresponding to the existence of a non-trivial intrinsic squeezed bispectrum can be probed by studying the GW power spectrum anisotropies. This is the case in particular for the STT correlator. Our analysis takes a step further and cross-correlates STT-sourced GW anisotropies with CMB temperature fluctuations. We find that for sizeable primordial non-Gaussianities, $F_{\rm NL}\gtrsim 10^3$, the intrinsic contribution dominates GW anisotropies and can be tested via cross-correlation with the CMB. The corresponding statistical error is of the order of a few percent. We conclude that large correlations between the SGWB and CMB  would be strongly suggestive of a primordial origin for the SGWB anisotropies. Naturally, the effectiveness of cross-correlations as a probe of primordial physics relies in no small part on the ever-increasing sensitivity of GW interferometers and depends on their frequency band. For example, a midband experiment such as BBO/DECIGO would improve detectability with respect to e.g. LISA in two ways: (i) it has a higher sensitivity to the SGWB and (ii) for a blue tensor spectral index, the signal is expected to be larger at those scales.

Our findings call for further investigation into several complementary directions. It would be interesting to perform a detailed derivation of the observables described in Section \ref{sec5}-B and \ref{sec5}-C \cite{future}. From a broader perspective, possible degeneracies with cross-correlations of  astrophysical origin should be accounted for. In particular, one may explore  correlations of the GW signal due, for example, to white dwarf binaries with CMB foregrounds (e.g. synchrotron emission, cosmic infrared background, or Sunyaev-Zel'dovich). While this goes beyond the scope of the present analysis, we hope to address this topic in a forthcoming work.   

\acknowledgements
We are delighted to thank Ameek Malhotra for insightful input on Section \ref{sec4}.
The work of PA was supported by the US Department of Energy through grant DE-SC0015655. NA acknowledges support by the University of Waterloo, Natural Sciences and Engineering Research Council of Canada, and the Perimeter Institute for Theoretical Physics (PITP). Research at PITP is supported in part by the Government of Canada through the Department of Innovation, Science and Economic Development Canada and by the Province of Ontario through the Ministry of Colleges and Universities. The work of MF is supported in part by the UK STFC grant ST/S000550/1 and the ``Atracci\'{o}n de Talento'' grant 2019-T1/TIC-15784. EAL is supported by an STFC AGP-AT grant
(ST/P000606/1). The work of GT is partially funded by STFC grant ST/P00055X/1. PA, ED, and MF acknowledge the hospitality of the Kavli Institute for Theoretical Physics, which is supported in part by the National Science Foundation under Grant No.\ NSF-PHY-1748958. We thank the organizers of the ``Gravitational Waves from the Early Universe'' workshop at NORDITA for their hospitality where some of this work was done. 

\bibliography{letter}

\end{document}